\documentclass[useAMS,usenatbib]{mn2e}
\usepackage{epsfig}
\usepackage{graphicx}
\usepackage{hyperref}
\begin{document}

\title[Statistics of Extinct Radio Pulsars]
{Statistics of Extinct Radio Pulsars}

\author[V.S. Beskin, S.A. Eliseeva]{V.S. Beskin$^1$ and S.A. Eliseeva$^2$
\thanks{E-mail: sa\_eliseeva@comcast.net; beskin@lpi.ru}\\
\thanks{Astronomy Letters, 2005, {\bf 31}, (4), 263-270, Translated from Russian by V. Astakhov}\\
$^1$ P.N.~Lebedev Physical Institute, Leninsky prosp. 53, Moscow, 119991, Russia
\\
$^2$ Moscow Institute of Physics and Technology, Dolgoprudny, Moscow region, 141700, Russia}

\pagerange{\pageref{firstpage}--\pageref{lastpage}} \pubyear{2002}

\maketitle

\label{firstpage}

\begin{abstract}
We study the statistical distribution of extinct radio pulsars -- isolated neutron stars, that just crossed the death line. An important element that distinguishes our study from other works is a consistent allowance for the evolution of the angle between the magnetic and spin axes. We find the distribution of extinct radio pulsars in spin period for two models: the model with hindered particle escape from the neutron star surface and the model with free particle escape. The total number of extinct radio pulsars is shown to be much smaller than for the model in which the evolution of an inclination angle is disregarded. The decrease in total number of extinct radio pulsars arises from the fact, that account for evolution of the inclination angle allows the transition of a neutron star to the propeller stage appear at much smaller periods ($P \sim 5-10$ s) than it is commonly assumed.
\end{abstract}

\section{Introduction}
\begin{figure}
\centering
\includegraphics[width=0.45\textwidth,keepaspectratio]{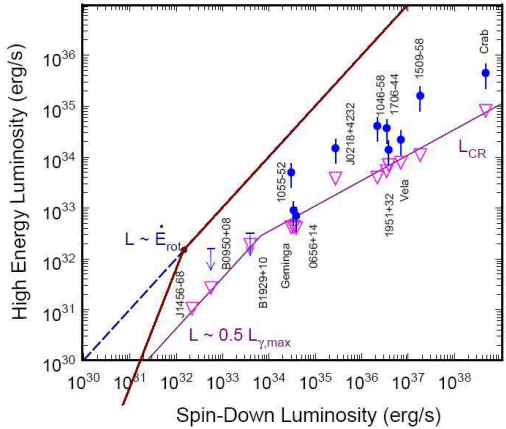}
\caption{\small{ Predicted and observed dependencies of the gamma-ray intensity for radio pulsars on the rate of rotational energy losses under the assumption of free particle escape from the neutron star surface. Red line presents extinct radio pulsars.}}
\label{fig1}
\end{figure}

Previously, we showed \citep{BE2003} that in the case of free
particle escape from the surface of an extinct radio pulsar for
small distances $d \sim 100-200 \mbox{ pc}$ typical of the nearest
neutron stars, their gamma-ray emission could be detected by the GLAST and AGILE missions.
Fig.~\ref{fig1} shows that our estimate of the intensity of the emission from extinct radio pulsars
for free particle escape from the neutron star surface agrees with
the estimate obtained by \citep{HM2002} for radio-emitting pulsars to
the numerical factor. The point is that the high-frequency emission
is associated mainly with primary particles and, therefore, it can
also take place when no secondary plasma is produced.

The mismatch between our estimate and the one done by \citep{HM2002}
is actually the gravitational factor. This difference arises due to
the fact that we consider particle acceleration to be effective only
at some distance from the star surface, at which the gravitational
effects can be neglected \citep{BE2003}. Let us draw provide more detailed consideration to this problem.

Here we consider the model with free particle escape from the
neutron star surface. Within this model the charge density of the particles flowing away from the polar cap differs only slightly from the Goldreich-Julian charge density. General relativity effects become significant near the neutron star surface (\citealt{MT1990}; \citealt{Beskin1990}). Since no particles are produced, the model of \citet{M1999}, in which no secondary plasma is generated, is appropriate in this case. The other model \citep{Arons1981} cannot be realized for extinct radio pulsars because it requires a reverse flow of secondary particles (for more details, see \citealt{Beskin1999}). Here a short explanation should be done. The Mestel's model presumes that none of the magnetic field lines near the pulsar surface are preferred because of the effects of general relativity. Therefore the particles cannot be accelerated in this region \citep{Beskin1990}. In other words, the longitudinal electric field that arises from the mismatch between the plasma charge density and the Goldreich charge density decelerates rather than accelerates plasma particles.

Nevertheless, the absence of regular acceleration does not imply that the plasma cannot fill the polar regions. Only effective particle acceleration is impossible. On the other hand, even at small distances from the pulsar surface $r > r_0$, where
\begin{equation}
r_0 \approx 1,8 \cdot 10^6 \left(\frac{M}{M_{\odot}}\right) P^{1/7}
\mbox{ cm},
\label{11}
\end{equation}

\noindent the general relativity effects become negligible. At such distances the difference between the plasma charge density and the Goldreich charge density results in particle acceleration, at least on the half of the polar cap where the magnetic field lines deviate from the spin axis (i.e., those magnetic field lines for which the angle between the magnetic field line and the spin axis increases with distance from the neutron star). Below, we assume that particle acceleration begins only after the radius $r = r_0$ is reached. At smaller radii plasma on open magnetic field lines rotates rigidly with a neutron star.

\renewcommand{\baselinestretch}{1.0}

\renewcommand{\baselinestretch}{1.5}
Now let us come back to the main point of this paper -- the
statistics of extinct radio pulsars. Since the intensity of the
emission strongly depends on the spin period of the neutron star,
radio-quiet pulsars with long spin periods (i.e., those that have
long ceased to radiate in the radio frequency range or those with
comparatively long spin periods at the stage of radio emission) can
no longer be detected even under the assumption of free particle
escape from the neutron star surface. Therefore, we can formulate
three necessary conditions under which extinct radio pulsars are
detectable: the validity of the model with free particle escape from
the neutron star surface, a relatively small distance to such stars,
and short spin period of a neutron star.

Clearly, the observability of extinct radio pulsars depends on their
total number in the Galaxy. Knowing the total number of extinct radio pulsars and assuming the equiprobable distribution of these stars in the Galaxy, we can estimate the number of extinct radio
pulsars can be within about 200 pc of the Sun. Another no
less important parameter is the spin period of a neutron star at
which a star passes to the stage of an extinct radio pulsar.

In this paper we consider the stationary distribution of extinct radio pulsars within the scope of two models: the model with hindered particle escape from the neutron star surface \citep{RS1975} and the model with free particle escape \citep{Arons1979}. An important element that distinguishes our study from other results is a consistent allowance for the evolution of an angle between the magnetic axis and the spin axis. Throughout the paper we assume that no neutron stars are born at this stage. Finally, we disregard the possibility of the magnetic field decay during the stages of active and extinct radio pulsars.

As a result, we find the distribution of extinct radio pulsars in spin period. We show that total number of extinct radio pulsars is much smaller than the number of extinct radio pulsars within the scope of model without the evolution of the inclination angle. The reason for such a decrease in the total number of these objects is the following: if the evolution of an inclination angle is taken into account, the transition to the propeller stage occurs at much shorter spin periods ($P \sim 5-10$ s) than it is possible for the standard model ($P \sim 100$ s), for which the evolution of an inclination angle is commonly disregarded.

\section{Basic equations}
\renewcommand{\baselinestretch}{1.0}

\begin{figure}
\centering
\includegraphics[width=0.45\textwidth,keepaspectratio]{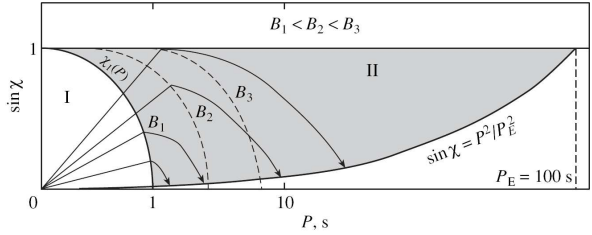}
\caption{\small{Isolated neutron star evolution under the assumption of hindered particle escape. Regions I and II correspond to active and extinct radio pulsars, respectively. The death line of active radio pulsars is shown for three different magnetic field strengths.}}
\label{fig2}
\end{figure}

\begin{figure}
\centering
\includegraphics[width=0.45\textwidth,keepaspectratio]{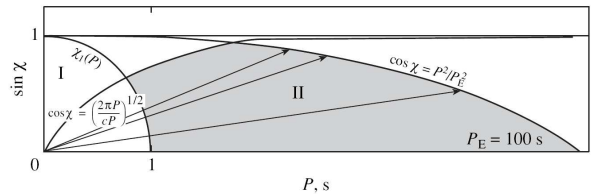}
\caption{\small{Isolated neutron star evolution under the assumption of free particle escape. Regions I and II correspond to active and extinct radio pulsars, respectively.}}
\label{fig3}
\end{figure}

\renewcommand{\baselinestretch}{1.5}
Our goal is to determine the number of neutron stars that are no longer radio pulsars, but have not yet passed to the propeller stage. In other words, we seek the distribution of neutron stars in the interval $P_d(\chi,B)<P<P_{pr}(\chi,B)$, where the period $P_d(\chi,B)$ corresponds to the death line of radio pulsars, and the period $P_{pr}(\chi,B)$ corresponds to the transition of neutron stars to the propeller stage (see Fig.~\ref{fig2} and \ref{fig3}). We take into account the fact that both $P=P_d(\chi,B)$ and $P=P_{pr}(\chi,B)$ depend on the angle of axial inclination $\chi$ (this is a new element that has not been studied previously).

The death line of radio pulsars can be written as follows \citep{BGI1993}
\begin{equation}
P_d(\chi) = B_{12}^{8/15} (\cos\chi)^{0.29} \mbox{ s}.
\label{deathline}
\end{equation}

It is known that the transition to the propeller stage occurs when the Schwarzman radius (i.e., the radius at which the pressure of the ambient medium is equal to the magnetodipole radiation pressure) is the same as the radius of the Alfven surface \cite{LPP1996}. Therefore, to determine the $\chi$-dependence of the transition period $P_{pr}$ we note that the magnetic dipole losses
\begin{equation}
L_{md} = -J_r \Omega {\dot\Omega}
= \frac{1}{6} \, \frac{B_0^2 \Omega^4 R^6}{c^3}
\sin^2\chi
\label{WW}
\end{equation}

\noindent ($B_0$ is the magnetic field at the pole of a neutron star with a radius $R$ and a moment of inertia $J_r$) are proportional to the combination $B_0 sin  \chi$. As a result, we obtain
\begin{equation}
P_{pr}(\chi) = P_{E} \sin^{1/2}\chi,
\label{prop}
\end{equation}

\noindent where the expression for $P_E$ has the standard form \citep{LPP1996}
\begin{eqnarray}
P_{E} \approx \frac{R}{c} \, \left(\frac{Rc^2}{GM}\right)^{1/2}
\, \frac{v_{\infty}}{c} \,
\left(\frac{B_0^2}{4\pi \rho_{\infty}v_{\infty}^2}\right)^{1/4}
\, \left(\frac{c_{\infty}}{v_{\infty}}\right)^{1/2} \\
\nonumber \approx 10^2 \, \frac{\mu_{30}^{1/2} c_{7}^{1/2} v_{7}^{1/2}}
{(B_{ext})_{-6}^{1/2}}  \mbox{  s}.
\label{PE}
\end{eqnarray}

\noindent Here, $\rho_{\infty}$ and $c_{\infty}$ are the density and the sound speed in the interstellar medium, respectively ($c_7$ is in units of $10^7$ cm s$^{-1}$, $B_{ext}^2=8 \pi \rho_{\infty}c_{\infty}^2$ is the corresponding energy density of the magnetic field ($(B_{ext})_{-6}$ is in units of $10^{-6}$ G). We also assume that the velocity of neutron stars $v_{\infty}$ is much greater than the sound speed in the interstellar medium \citep{PCT2000}.

To determine the stationary distribution function of extinct radio pulsars, we introduce their distribution function $N(P,\chi,B)$ in spin period $P$, inclination angle $\chi$, and magnetic field $B$. The kinetic equation can be presented in general form as \citep{BGI1993}
\begin{equation}
\frac{\partial}{\partial P}
\left(\frac{\delta P}{\delta t}N\right)
+\frac{\partial}{\partial \chi}
 \left(\frac{\delta \chi}{\delta t}N\right)=U.
\label{kin}
\end{equation}

In other words, we ignore the magnetic field decay over the lifetime of an extinct radio pulsar the same way as we did for radio pulsars. The validity of the assumption for normal radio pulsars seems proven, since their lifetime does not exceed the characteristic magnetic-field evolution time \citep{PP2000}. As we show below, if the evolution of the angle of inclination is taken into account, the lifetime of a neutron star during the extinct radio pulsar stage is also short enough.

The derivatives $\delta P/\delta t$ and $\delta\chi/\delta t$ (\ref{kin}) are known functions of the magnetic field $B$, the period $P$, and the angle of inclination $\chi$. Further, if hindered particle escape takes place and the spin period of a pulsar $P>P_d(\chi,B)$, plasma fills only the inner region of the pulsar magnetosphere \citep{IM1995}. Thus, for the model with hindered particle escape from the radio-pulsar surface, we assume that the energy losses closely match the magnetic dipole losses (\ref{WW}):
\begin{eqnarray}
\frac{\delta P}{\delta t}=\frac{2\pi^2}{3} \, \frac{B_0^2R^6}{J_r P c^3}
\approx -10^{-15}
B_{12}^{2}P^{-1}\sin^{2}\chi,
\label{deriv_hind1}
\\
\frac{\delta \chi}{\delta t} \approx -10^{-15} B_{12}^{2}P^{-2}\sin\chi\cos\chi \mbox{ s}^{-1}. \hspace{3em}
\end{eqnarray}

It should be emphasized here that within the scope of hindered particle escape model the following ratio is invariant for each individual neutron star throughout the stage of extinct radio pulsars.
\begin{equation}
I_d=\frac{\cos\chi}{P}
\label{id}
\end{equation}

\noindent As a result, an inclination angle of each extinct radio pulsar decreases during the evolution (providing the validity of hindered particle escape model). Moreover, an inclination angle $chi$ decreases at the same rate as the spin period of the neutron star increases.

As for free particle escape model, here we assume that the neutron star magnetosphere is still completely filled with plasma that screens the longitudinal electric field. In this case, the magnetic dipole losses are known to be completely screened. Therefore, all the energy losses must be associated with the longitudinal currents in the magnetosphere (\citealt{BGI1993}, \citealt{M1999}). In other words, the energy losses are the same as those for radio pulsars. Thus, within the scope of free particle escape model we obtain \citep{BGI1993}
\begin{eqnarray}
\frac{\delta P}{\delta t} \approx 10^{-15}B_{12}^{10/7}P^{1/14}\cos^{2d}\chi, \hspace{2.5em} 
\label{deriv_hind2}
\\
\frac{\delta \chi}{\delta t} \approx 10^{-15} B_{12}^{10/7}P^{1/14}\cos^{2d-1}\chi \mbox{ s}^{-1}.
\end{eqnarray}

\noindent Here $2d \simeq 1.5$ is a dimensionless parameter. So, for the model with free particle escape the ratio
\begin{equation}
I_{\rm c} = \frac{\sin\chi}{P}
\label{ic}
\end{equation}

\noindent is invariant throughout the time a neutron star spends as an extinct radio pulsar. As we can see, both an inclination angle $\chi$ and a spin period increase at the same rate providing the particles freely escape a neutron star surface. An inclination angle for this model tends to $90^{\circ}$. So, for both models of particle escape function the angles tend to pass into the region where the energy losses are minimumal.

Further, function $U$ on the right-hand side of Eq. (\ref{kin}) is the source of extinct radio pulsars. To make a reasonable assumption about this function we should recall how it is defined in case of active radio pulsars. Using the results obtained in the monograph \cite{BGI1993}, we assume that the birth function of neutron stars can be presented as a product of the probabilities of their initial distributions in angle $\chi$, period $P$, and magnetic field $B$:
\begin{equation}
U = U(\chi)U_P(P)U_B(B).
\label{source1}
\end{equation}

Unfortunately, we have no information about the distribution of nascent neutron stars in axial inclination $\chi$ \citep{TM1999}. Therefore, we have to assume that the distribution of neutron stars in their initial inclination angles $\chi$ is equiprobable, so that
\begin{equation}
U(\chi) = \frac{2}{\pi}.
\label{source}
\end{equation}

In order to determine the birth function $U(P)$, we find the period distribution for pulsars with characteristic ages $\tau_{\rm D} =P/ 2 \dot P$ greater than the kinetic ages $\tau_{\rm kin} = z/v_{\perp}$, which can be determined for pulsars with known proper motions. Here $v_{\perp}$ is the pulsar velocity perpendicular to the Galactic plane. We use the fact that the thickness of the disc of radio pulsars, $z_p \approx 300$ pc, is much larger than the thickness of the disc of supernova remnants, $z_{sn} \approx 10$ pc \cite{MS1984}, where most of the neutron stars are assumed to be born. As a result, the radio pulsars from this sample have not changed their periods significantly. Therefore, the distribution of such pulsars in spin period $P$ should be close to the true birth function of neutron stars $U(P)$.

\begin{figure}
\centering
\includegraphics[width=0.45\textwidth,keepaspectratio]{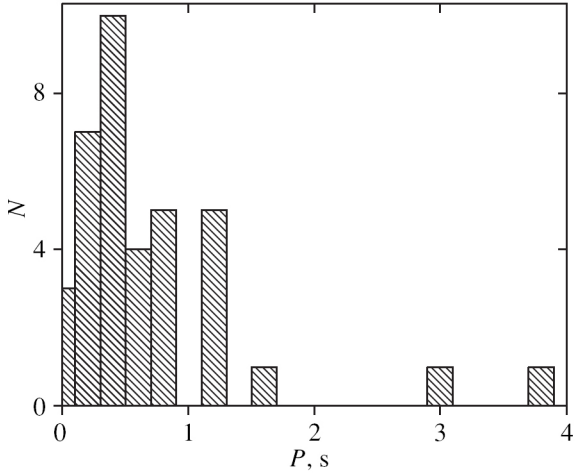}
\caption{\small{Spin period distribution for 37 active radio pulsars with $\tau_D > \tau_{kin}$. Data from the ATNF catalogue are
used to obtain the distribution.}}
\label{fig4}
\end{figure}
\renewcommand{\baselinestretch}{1.5}
 Fig.~\ref{fig4} generally matches the distribution obtained previously \citep{BGI1993}. The distribution seems to be nearly equiprobable. At least, it has no distinct maximum at short periods. Therefore, in case of active radio pulsars we can assume that
\begin{equation}
U_P(P)= U_P = \mbox{const}.
\label{source_UP}
\end{equation}

We use this distribution function to analyze the statistical distribution of radio pulsars. On the other hand, we see that a sharp (almost to zero) decrease in the number of radio pulsars is observed at periods $P>1$ s. At the same time, the decrease in the number of radio pulsars for the complete sample in the period interval $1<P<2$ s is not so essential. Therefore, we assume that no neutron stars are born at the stage of extinct radio pulsars: $U=0$.

Finally, magnetic-field distribution function of radio pulsars can be written as \citep{bgi1993}
\begin{equation}
U_B(B) = \frac{1}{B_0\Gamma(\gamma+1)}
\left(\frac{B}{B_0}\right)^{\gamma}\exp(-B/B_0),
\label{source_UB}
\end{equation}

\noindent where $B_0=10^{12}$ G and $\gamma=2$. In order to refine the magnetic field dependence of the distribution function of radio pulsars, we use the logarithmic chart plotting the number of radio pulsars $N(B>B_0)$ vs. magnetic field $B$ for 1348 pulsars \citep{MCatalog}. The pulsars are divided into two groups by the dimensionless parameter $Q$ that reflects their basic characteristics:
\begin{equation}
Q=2\left(\frac{P}{1 \mbox{ s}}\right)^{11/10}
\left(\frac{\dot P}{10^{-15}}\right)^{-4/10}.
\label{Q}
\end{equation}
\renewcommand{\baselinestretch}{1.0}

\begin{figure}
\centering
\includegraphics[width=0.45\textwidth,keepaspectratio]{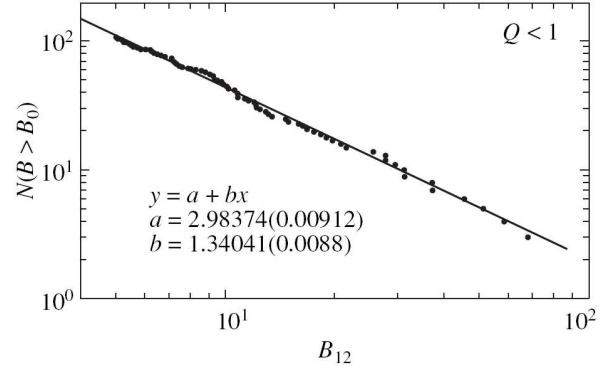}
\caption{\small{Distribution of young radio pulsars ($Q<1$) in magnetic field $B_{12}$ on a logarithmic scale. Data from the ATNF catalogue are used to construct the distribution.}}
\label{fig5}
\end{figure}

\begin{figure}
\centering
\includegraphics[width=0.45\textwidth,keepaspectratio]{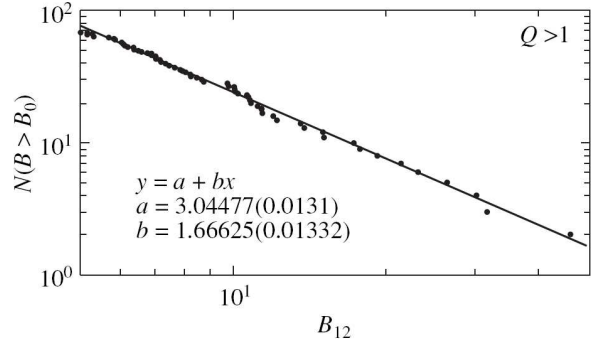}
\caption{\small{Distribution of old radio pulsars ($Q>1$) in magnetic field $B_{12}$ on a logarithmic scale. Data from the ATNF catalog are used to construct the distribution.}}
\label{fig6}
\end{figure}

\renewcommand{\baselinestretch}{1.5}

The pulsars with $Q>1$ and $Q<1$ lie near and far from the death line (for more details, see \citealt{BGI1984}). Figures~\ref{fig5} and \ref{fig6} show the distribution functions of extinct radio pulsars with $Q<1$ and $Q>1$, respectively. Surprisingly, despite the fact that the number of observed radio pulsars increased more than 3 times over the last ten years, we obtained almost the same results that were found previously for the statistics of about 400 pulsars. As a result, the distribution function of radio pulsars with $Q<1$ can be written as follows \cite{BGI1993}
\begin{eqnarray}
N_{1}(P,\chi;B_{0})
=2k_{N}N_{f}P\frac{1}{B_{k}}\left(\frac{B_{0}}{B_{k}}\right)^{0.57}
\left(1+\frac{B_{0}}{B_{k}}\right)^{-3.7} \times \\
\nonumber \times
\frac{1-\cos{\chi}}{\sin^{2}{\chi}}\cos^{-0.5}{\chi} \times \Theta.\hspace{6.5em} 
\label{N1}
\end{eqnarray}

\noindent Here, $k_N \approx 4.4$ is a normalizing factor, $\Theta = \theta[P_1(\chi)-P]$, and $N_f$ is a number of active radio pulsars far from the death line.

Thus, we consider the stationary distribution of extinct radio pulsars without neutron stars being born at this stage and with the magnetic-field evolution disregarded. Accordingly, the kinetic equation (\ref{kin}) can be rewritten as
\begin{equation}
\frac{\partial}{\partial P}\left(N\frac{dP}{dt}\right)
+\frac{\partial}{\partial \chi}\left(N \frac{d\chi}{dt}\right)=0.
\label{kin_stat}
\end{equation}

\noindent The solution of the kinetic equation for the model with hindered particle escape is given by
\begin{equation}
N(P, \chi)=\frac{F(I_d)}{\sin \chi}.
\label{N_hind}
\end{equation}

\noindent For the model with free particle escape from the pulsar surface it is
\begin{equation}
N(P, \chi)=\frac{F(I_c)}{\cos \chi}.
\label{N_free}
\end{equation}

\noindent To solve the kinetic equation for the two models, we should define the boundaries of the region of extinct radio pulsars. For the model with hindered particle escape from the neutron star surface (the Ruderman-Sutherland model) these boundaries are determined by the death line of radio pulsars ($\Theta_1=\theta[P_1(B, \chi)-P]$) and by the line where the transition of extinct radio pulsars to the propeller stage takes place ($\Theta_2=\theta[sin\chi - P^2/P_E^2]$). In within the scope of free particle escape model, the transition to the region in which an inclination angle is close to $90^{\circ}$ \citep{BN2004} should be added to the above boundaries.

Let us designate the regions of active and extinct radio pulsars as regions I and II, respectively. To determine the function $F(I)$, we should match the solutions at the boundary between the regions of active (region I) and extinct (region II) radio pulsars. Due to the fact that the neutron star fluxes on both sides of the death line should remain equal, we obtain for the model with hindered particle escape
\begin{equation}
N_{2}[P_{d}(\chi),\chi]=\frac{{\displaystyle{\dot P_1
\left(-\frac{d \chi_1}{dP}
-\frac{\partial I_{c} /\partial P}{\partial I_{c} /\partial
\chi }\right)}}}{{\displaystyle{\dot P_{2}\left(-\frac{{d}
\chi_1}{{d}P}-\frac{\partial I_{d}
/\partial P}{\partial I_{d}/\partial
\chi}\right)}}}N_1[P_{d}(\chi), \chi],
\label{N2_hind}
\end{equation}

\noindent Here the period and angle derivatives correspond to the death line of pulsars $P=P_d(\chi)$. At the same time, for the model with free particle escape, we obtain
\begin{equation}
N_{2}[P_d(\chi),\chi]=N_1[P_d(\chi), \chi].
\label{N2_free}
\end{equation}

As a result, in the Ruderman-Sutherland model, we can approximate the function $F(I)$ by the following expression
\begin{equation}
F(I_d) \approx A(B_{12}) \frac{\cos^n \chi}{P^n},
\label{F_hind}
\end{equation}

\noindent where $n \approx -0.6$, and $A(B_{12}) \propto B_{12}(1 + B_{12})^{-3.7}$.

\section{Ruderman-Sutherland model}
Let us first consider the statistics of extinct radio pulsars assuming the model of hindered particle escape from the neutron star surface is valid. As it was mentioned above, the particle work function is fairly large in this case, and particles fill only the equatorial regions of the magnetosphere with the strongly curved magnetic field, in which secondary particles can still be produced. Therefore, the energy losses for the Ruderman-Sutherland model can be described with a sufficient accuracy by a magnetic dipole mechanism.

Let us now turn to Eq. (\ref{N2_hind}) derived from the conservation condition for the neutron star flux on the death line of radio pulsars. This expression defines the function $F(I)$ and, hence, the distribution of extinct radio pulsars in period $P$, magnetic field $B$, and inclination angle $\chi$ for the model in question. In this equation, the derivative $\dot P_1$ corresponds to the region of active radio pulsars and can be specified by \cite{BGI1993}:
\begin{equation}
\frac{\delta P}{\delta t} \approx 10^{-15}B_{12}^{10/7}P^{1/14}\cos^{1.5}\chi.
\label{deriv_hind_norm}
\end{equation}

At the same time, the derivative $\dot P_2$ is specified in the region of extinct radio pulsars, and Eq. (\ref{deriv_hind1}) should be used here. Next, let us recall that the region of extinct radio pulsars in the scope of hindered particle escape model is limited by the death line of active radio pulsars (\ref{deathline}) and the line of the transition to the propeller stage (\ref{prop}).

As a result of the above expressions and using the fact that distribution function for normal radio pulsars is defined by Eq. (\ref{N1}), Eq. (\ref{N2_hind}) can be rewritten as
\begin{eqnarray}
N_2[P_d(\chi), \chi]=0.7 k_N N_f B_{12}^{8/7}(1+B_{12})^{-3.7} \times \hspace{0.3em}\\
 \nonumber \times \frac{1-cos \chi}{sin^4 \chi} cos^{1.7} \chi (3.5+tg^2 \chi).
\label{N2_hind1}
\end{eqnarray}

On the other hand, the solution of the kinetic equation for the model with hindered particle escape providing the distribution of extinct radio pulsars is stationary, no stars are born at this stage, and the magnetic-field evolution is disregarded, can be represented by Eq. (\ref{N_hind}). Thus, for the distribution of extinct radio pulsars on the death line, we obtain
\begin{equation}
N_2[P_d(\chi), \chi]=\frac{F(I_d)}{sin \chi}=\frac{F(cos^{0.71}\chi/B_{12}^{8/15})}{sin \chi}.
\label{N2_hind2}
\end{equation}

Further, we rewrite the distribution function for the extinct radio pulsars derived from the conservation condition for the flux through the death in the same form as Eq.~\ref{N2_hind2}. For this purpose, we use the following change of variable:
\begin{equation}
\xi=\frac{cos^{0.71} \chi}{B_{12}^{8/15}}.
\label{xi}
\end{equation}

As a result, the distribution function for the extinct radio pulsars can be finally transformed to
\begin{eqnarray}
N_2[P, B, \chi]= 0.7 k_N N_f B_{12}^{8/7}(1+B_{12})^{-3.7} \times \\
\nonumber \times \frac{G(cos \chi/P)}{sin \chi} \Theta_1 \Theta_2, \hspace{4.3em}
\label{N2_hind_result}
\end{eqnarray}

\noindent where
\begin{equation}
G(\xi)=\frac{1-c(\xi)}{s^3(\xi)}(3.5+\frac{s^2(\xi)}{c^2(\xi)})c^{1.7}(\xi).
\label{G}
\end{equation}

\noindent Here we use two auxiliary functions:
\begin{eqnarray}
c(\xi)=B_{12}^{0.75} \xi^{1.4}=cos \chi, \hspace{1.7em}
\label{c}
\\
s(\xi)=[1-c^2(\xi)]^{1/2}=sin \chi.
\label{s}
\end{eqnarray}
\renewcommand{\baselinestretch}{1.0}

\begin{figure}
\centering
\includegraphics[width=0.45\textwidth,keepaspectratio]{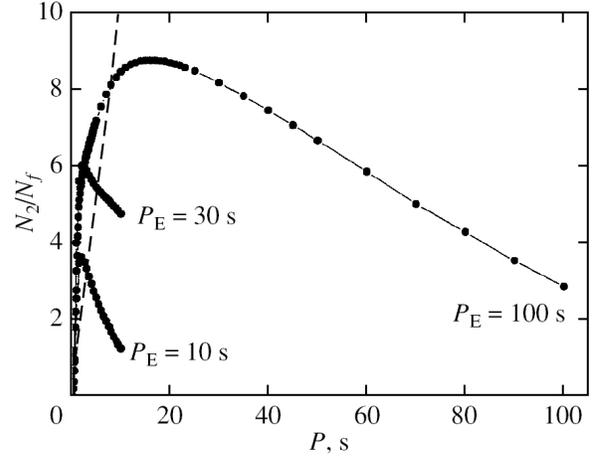}
\caption{\small{Distribution of extinct radio pulsars $N_2(P)$ in the scope of hindered particle escape model with (solid lines) and without (dashed line) allowance for the evolution of the inclination angle.}}
\label{fig7}
\end{figure}

\renewcommand{\baselinestretch}{1.5}
Figure~\ref{fig7} shows the distribution of extinct radio pulsars in spin period $P$ under the assumption of hindered particle escape for three different values of limiting spin period $P_E$ (\ref{PE}). The results of numerical calculations shown in this figure are in good agreement with the analytical estimate (\ref{F_hind}) obtained for the Ruderman-Sutherland model. As we see, if the evolution of an angle of inclination of the magnetic axis to the spin axis is consistently taken into account, total number of neutron stars at the extinct radio pulsar stage proves to be much smaller than the same total number within the model for which the evolution of an inclination angle is disregarded. On the other hand, the number of pulsars with comparatively short periods $P \sim 2-4$ s at this stage is even larger than that for the standard model (the dashed line in Fig.~\ref{fig7}). This is attributable to the fast transition of neutron stars to the region of small angles of inclination, where they are not only accumulated through the decrease in spin-down rate, but also disappear due to the transition to the propeller stage.

Finally, it is very important that, if the evolution of the angle of axial inclination is taken into account, the transition to the propeller stage can also occur at short neutron star spin periods $P \sim 5-10$ s, or even at $P \sim 1-2$ s (see Fig.~\ref{fig2}). Indeed, as we mentioned above, the transition to the propeller stage occurs when the Schwarzman radius, which can be determined from the balance between the magnetodipole and external radiation pressures, becomes equal to the Bondi-Hoyle radius. However, since the magnetodipole losses at small angles of axial inclination are essentially suppressed, the transition to the propeller stage takes place even at short neutron star spin periods:
\begin{equation}
P_{pr}=P_E (sin \chi)^{1/2} \approx P_E sin \chi,
\label{P_pr_hind}
\end{equation}

\noindent which is about $5-10$ s at $P_E \sim 100$ s (\ref{PE}). Recall once again that if the model with hindered particle escape from the neutron star surface is valid, extinct radio pulsars cannot be detected with currently available instruments.

\section{Arons model}
Let us now discuss the statistics of extinct radio pulsars assuming the model with free particle escape from the neutron star surface is valid. Recall that for this model, the energy losses for both extinct and active radio pulsars are the same and are associated with the longitudinal currents in the magnetosphere. Wa want to stress here that within the model of free particle escape gamma-ray emission from extinct radio pulsars with short (several seconds) spin periods and providing relatively small distances to such stars can be detected.

Let us consider Eq. (\ref{N2_free}), which is derived from the condition for the equality between the neutron star fluxes on both sides of the death line. Using the distribution function of active radio pulsars (\ref{N1}) derived previously \cite{BGI1993} and refined by using the new data in this paper, we obtain the following expression for the distribution function of extinct radio pulsars on the death line:
\begin{eqnarray}
N_2[P_d(\chi), \chi]=0.6 k_N N_f B_{12}^{1.1}(1+B_{12})^{-3.7} \times \\
\nonumber \times \frac{1-cos \chi}{sin^2 \chi}cos^{-0.21}\chi. \hspace{3.5em}
\label{N2_free1}
\end{eqnarray}

On the other hand, the solution of the kinetic equation for the model of free particle escape from the neutron star surface can be represented as (\ref{N_free}). Accordingly, for the pulsars on the death line, the solution of the kinetic equation is given by
\begin{equation}
N_2[P_d(\chi), \chi]=\frac{F(I_d)}{cos \chi}=\frac{F[sin \chi /(B_{12}^{8/15}cos^{0.29}\chi)]}{cos \chi}.
\label{N2_free2}
\end{equation}

Recall that, as for the Ruderman-Sutherland model, this solution is obtained assuming the distribution of extinct radio pulsars to be stationary. At the same time, we disregarded the possibility of stars being born at this stage and the possibility of magnetic field decay.

Next, it is necessary to consider the boundaries of the region of extinct radio pulsars for the model with free particle escape. This region is limited by the death line of radio pulsars, the line of the transition of pulsars to the propeller stage, and the line of the transition of pulsars to the region where the angle of inclination of the magnetic axis to the spin axis of a neutron star is close to $90^{\circ}$ (see Fig.~\ref{fig3}). As in the case of hindered particle escape, the death line of radio pulsars is given by Eq.(\ref{deathline}). The transition of extinct radio pulsars to the propeller stage is defined as follows:
\begin{equation}
P_{pr}(\chi)=P_E cos^{1/2} \chi.
\label{P_pr_free}
\end{equation}

Here, the expression for the limiting period $P_E$ also has the standard form (\ref{PE}). Finally, the third boundary of the region of extinct radio pulsars is the line of the transition to the region where an inclination angle approaches $90^{\circ}$. For this boundary we have the following expression \citep{BN2004}:
\begin{equation}
cos \chi=\left(\frac{2 \pi R}{cP}\right)^{1/2}.
\label{chi90}
\end{equation}

Unfortunately, within the scope of free particle escape  model, the distribution function of extinct radio pulsars derived from the flux conservation on the death line cannot be analytically rewritten in the same form as (\ref{N2_free2}). Still, we can find analytical approximations for the large or small inclination angles. Thus, for small angles of axial inclination,
\begin{equation}
F(I)=\frac{1}{\pi}k_N N_fB_{12}^{1.1}(1+B_{12})^{-3.7}.
\label{F_free_small}
\end{equation}

At the same time, for angles of inclination close to $90^{\circ}$, we obtain the following expression:
\begin{equation}
F(I)=\frac{2}{\pi}k_N N_f B_{12}^{0.2}(1+B_{12})^{-3.7}I^{-1.7}.
\label{F_free_large}
\end{equation}

Now, using the expressions obtained for large and small angles of inclination $\chi$, we can approximate the function $F(I)$ for all inclination angles $<0\chi<\pi/2$ as
\begin{equation}
F(I)=\frac{0.51}{(1+0.1B^{1.7}I^{3.1})^{0.9}}.
\label{F_free_result}
\end{equation}

Finally, the distribution of extinct radio pulsars for the model with free particle escape can be represented by the function
\begin{eqnarray}
N_2[P, B, \chi]=0.3k_N N_f \times \hspace{12em}\\
\nonumber \times \frac{B_{12}^{1.1}(1+B_{12})^{-3.7}}{cos \chi (1+0.1B^{1.7}(sin \chi/P)^{3.1})^{0.9}} \Theta_1...
\label{N2_free_result}
\end{eqnarray}

Recall that the normalizing factor in this expression is $k_N \approx 4.4$, and $N_f$ gives the number of normal radio pulsars far from the death line.
\renewcommand{\baselinestretch}{1.0}

\begin{figure}
\centering
\includegraphics[width=0.45\textwidth,keepaspectratio]{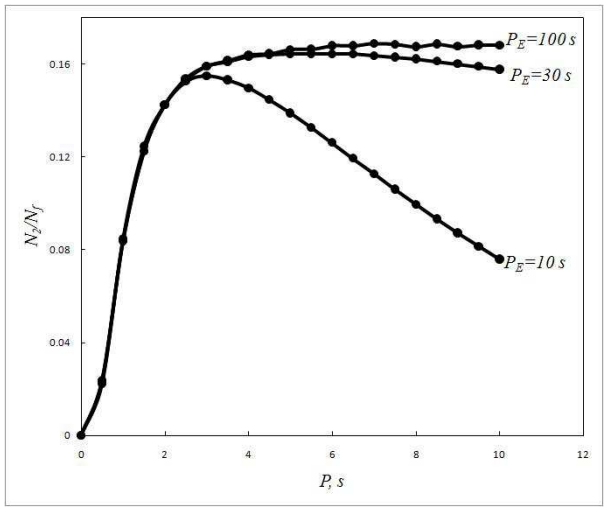}
\caption{\small{Distribution of extinct radio pulsars $N_2(P)$ within the scope of free particle escape model.}}
\label{fig8}
\end{figure}

\renewcommand{\baselinestretch}{1.5}
Figure~\ref{fig8} shows the distribution of extinct radio pulsars in spin period $P$ under the assumption of free particle escape from the neutron star surface. As we can see, for this model, just as for the Ruderman-Sutherland model, consistent allowance for the evolution of the angle between the magnetic axis and the spin axis causes the total number of extinct radio pulsars to decrease. Such a decrease appears because neutron stars can pass to the propeller stage at much shorter spin periods than assumed previously. Indeed, if inclination angles are close to $90^{\circ}$, extinct radio pulsars can pass to the propeller stage even at periods
\begin{equation}
P_{pr}=P_E (cos \chi)^{1/2}.
\label{P_pr_free90}
\end{equation}

Thus, if an angle of inclination of the magnetic axis to the spin axis is close to $90^{\circ}$, the spin period of a pulsar passing to the propeller stage could be $P \sim 5-10$ s.

\section{Conclusions}
We found the theoretical distribution of extinct radio pulsars in spin period within the scope of two models: the model with hindered particle escape from the neutron star surface and the model with free particle escape. In both cases, we consistently took into account the evolution of an angle between the magnetic axis and the spin axis.

As a result, we showed that for both models all the details of the calculations, such as the death line of active radio pulsars, the transition of extinct radio pulsars to the propeller stage, and the evolution of extinct radio pulsars, actually depend severely on the evolution of an inclination angle. Unfortunately, with a few exceptions (\citealt{BGI1993}, \citealt{RFP2001}), no consistent analysis of the evolution of an inclination angle has been performed previously.

As we proved, such an allowance for the evolution of the axial inclination in analyzing the statistics of extinct radio pulsars causes the total number of such radio pulsars to decrease, thereby making it difficult to detect them with currently available instrumentation (the GLAST or AGILE missions). The decrease is due to the possibility for extinct radio pulsars to pass to the propeller stage at shorter spin periods $P \sim 5-10$ s than it is possible for the model without account of inclination angle's evolution. Nevertheless, if we consider the model with hindered particle escape and take into account the evolution of an angle $\chi$, the number of extinct radio pulsars with particular short spin periods ($P \sim 2-4$) s is even larger than that in the standard model. Finally, it should also be stressed here that the detection of extinct radio pulsars would be direct evidence for free particle escape from the neutron star surface.

\section{Acknowledgements}
We are grateful to S.B.Popov and M.E.Prokhorov for fruitful discussions and numerous advice. This work was supported by the Russian Foundation for Basic Research (project nos. 02-02-16762 and 05-02-17700).


\begin{thebibliography}{30}

\bibitem[\protect\citeauthoryear{Arons}{1979}]{Arons1979}
Arons J., 1979, Space Sci. Rev., 24, 437

\bibitem[\protect\citeauthoryear{Arons}{1981}]{Arons1981}
Arons J., 1981, Astrophys. J., 248, 1099

\bibitem[\protect\citeauthoryear{Beskin}{1990}]{Beskin1990}
Beskin V.S, 1990, Pis'ma Astron. Zh., 16, 665 [Sov.
Astron. Lett. 16, 286 (1990)]

\bibitem[\protect\citeauthoryear{Beskin}{1999}]{Beskin1999}
Beskin V.S, 1999, Phys. Uspekhi  40, 659

\bibitem[\protect\citeauthoryear{Beskin \& Eliseeva}{2003}]{BE2003}
V.S. Beskin and S.A. Eliseeva, 2003, Astron. Lett., 29, 20

\bibitem[\protect\citeauthoryear{Beskin, Gurevich \& Istomin}{Beskin et al.}{1884}]{BGI1984}
Beskin V.S., Gurevich A.V., Istomin Ya.N., 1984, Astrophys. Space Sci., 102, 301

\bibitem[\protect\citeauthoryear{Beskin, Gurevich \& Istomin}{Beskin et al.}{1993}]{BGI1993}
Beskin V.S., Gurevich A.V., Istomin Ya.N., 1993,
''Physics of the pulsar magnetosphere'', Cambridge University Press

\bibitem[\protect\citeauthoryear{Beskin \& Nokhrina}{2004}]{BN2004}
V.S. Beskin and E.E. Nokhrina, 2004, Astron. Lett., 30, 685

\bibitem[\protect\citeauthoryear{Harding, Muslimov \& Zhang}{2002}]{HM2002}
A. K. Harding, A. G. Muslimov and B. Zhang, 2002, Astrophys. J., 576, 366

\bibitem[\protect\citeauthoryear{Istomin \& Mosyagin}{1995}]{IM1995}
Ya. N. Istomin and D. V. Mosyagin, 1995, Astron. Zh., 72, 826 [Astron. Rep. 39, 735 (1995)]

\bibitem[\protect\citeauthoryear{Lipunov}{1987}]{Lipunov1987}
V.M. Lipunov, 1987 ''Astrophysics of Neutron Stars'', Nauka,
Moscow

\bibitem[\protect\citeauthoryear{Lipunov, Postnov \& Prokhorov}{Lipunov et al.}{1996}]{LPP1996}
V.M. Lipunov, K.A. Postnov and M.E. Prokhorov, 1996, Astrophys.~Space~Phys., 9, 1

\bibitem[\protect\citeauthoryear{Manchester et al.}{2005}]{MCatalog}
Manchester R.N., Hobbs G.B., Teoh A., Hobbs M., 2005, AJ, 129, 1993

\bibitem[\protect\citeauthoryear{Marochnik \& Suchkov}{1996}]{MS1984} Marochnik L.S., Suchkov A.A., 1996, ''The
Milky Way Galaxy'', Gordon and Breach

\bibitem[\protect\citeauthoryear{Mestel}{1999}]{M1999}
Mestel L., 1999, ''Stellar Magnetism'', Oxford University Press

\bibitem[\protect\citeauthoryear{Muslimov \& Tsygan}{1990}]{MT1990}
A.G. Muslimov and A.I. Tsygan, Astron. Zh., 1990, 67, 263 [Sov. Astron. 34, 133 (1990)].

\bibitem[\protect\citeauthoryear{Popov, Colpi \& Treves}{Popov et al.}{2000}]{PCT2000}
S.B. Popov, M. Colpi and A. Treves, 2000, Astrophys. J., 530, 896

\bibitem[\protect\citeauthoryear{Popov \& Prokhorov}{2000}]{PP2000}
S.B. Popov and M.E. Prokhorov, 2000, Astron. Astrophys., 357, 164

\bibitem[\protect\citeauthoryear{Regimbau \& de Freitas Pacheco}{2001}]{RFP2001}
Regimbau T., de Freitas Pacheco J.A., 2001, A\&A, 374, 182

\bibitem[\protect\citeauthoryear{Ruderman \& Sutherland}{1975}]{RS1975}
Ruderman M.A., Sutherland P.G., 1975, ApJ, 196, 51

\bibitem[\protect\citeauthoryear{Tauris \& Manchester}{1998}]{TM1998}
Tauris T.M., Manchester R.N., 1998, MNRAS, 298, 625

\end{thebibliography}
\end{document}